\documentclass[11pt]{cernrepp}
\usepackage{graphicx}
\usepackage{here}
\usepackage{epsfig}
\usepackage{cite}

\begin{document}
 \title{CONFIDENCE LIMITS:  WHAT IS THE PROBLEM?
         IS THERE \\{\it THE} SOLUTION?}
\author{G. D'Agostini}
\institute{ Universit{\`a} ``La Sapienza'' and
        Sezione INFN di Roma 1, Rome , Italy, and CERN, Geneva, Switzerland\\
        {\rm Email}: {\tt giulio.dagostini@roma1.infn.it}\\ 
        {\rm URL}: {\tt http://www-zeus.roma1.infn.it/$^\sim$agostini}}
\maketitle
\begin{abstract}
This contribution to the debate on confidence limits
focuses mostly on the case of measurements with `open likelihood', in the
sense that it is defined in the text. 
I will show that, though a prior-free assessment 
of {\it confidence} is, in general,
not possible, still a search result can be reported
in a mostly unbiased and efficient way, which satisfies 
some desiderata which I believe are shared by the people
interested in the subject. The simpler case of `closed likelihood'
will also be treated, and I will discuss why a uniform prior 
on a sensible quantity is
a very reasonable choice for most applications. 
In both cases, I think that 
much clarity will be achieved if we remove from  scientific 
parlance the misleading expressions 
 `confidence intervals' and `confidence levels'.
\end{abstract}

\begin{flushright}
{\sl \small ``You see, a question has arisen,}\\
{\sl \small about which we cannot come to an agreement,}\\
{\sl \small probably because we have read too many books''}\\
(Brecht's Galileo)\footnote{{\it ``Sehen Sie, es  ist eine Frage
enstanden, \"uber die wir uns nicht einig werden k\"onnen, 
wahrscheinlich, weil wir zu viele B\"ucher gelesen 
haben.''} (Bertolt Brecht, {\sl Leben des Galilei}).}
\end{flushright}

\section{INTRODUCTION}

The blooming of papers on `limits' in the past couple of
years~\cite{cl_papers,FC,LEP,PDG,Zech,ci,Higgs,
Ciampolillo,Eitel,noi,Punzi}
and a 
workshop\cite{CLW} entirely dedicated to the subject
are striking indicators of the level of the problem. 
It is difficult not to agree that at the root of the problem 
is the standard physicist's education in statistics, 
based on the collection of frequentistic prescriptions, 
given the lofty name of  
`classical statistical theory' by the their supporters, 
`frequentistic adhoc-eries'\footnote{For example, 
even Sir Ronald Fisher used
to refer to Neyman's statistical confidence method as 
``that technological and commercial apparatus  which is known as an
acceptance procedure''~\cite{Fisher}. In my opinion,  
the term  `classical' is misleading, as are the results of 
these methods. The name gives the impression of being  analogous
to `classical physics', which was developed by our `classicals',
and that still holds for ordinary problems. Instead, the
classicals of probability theory, like Laplace, Gauss, Bayes, 
Bernoulli and Poisson, had an approach to the problem more 
similar to what we would call nowadays `Bayesian'
(for an historical account see Ref.~\cite{Hald}).}
by their opponents. 
In fact, while in routine measurements characterised by 
a narrow likelihood `correct numbers' are obtained by 
frequentistic prescriptions 
(though the intuitive interpretation that physicists attribute to them  
is that of probabilistic 
statements\footnote{It is a matter of fact~\cite{maxent98} 
that confidence levels are 
intuitively thought of (and usually taught) 
by the large majority of physicists
as degrees of belief on true values, 
although the expression `degree of belief' is 
avoided, because ``beliefs are not scientific''. Even 
books which do insist on stating that probability statements
are not referred to true values (``true values are constants 
of unknown value'')
have a hard time explaining the real meaning
of the result, i.e. something which maps into the human mind's  
perception of uncertain events. So, they are forced to use
ambiguous sentences which remain stamped in the memory of 
the reader much more than the frequentistically-correct twisted
reasoning that they try to explain. For example a classical 
particle physics statistics book~\cite{Frodesen} speaks about 
``the faith we attach to this statement'', as if `faith' 
was not the same as degree of belief. 
Another one~\cite{Eadie} introduces the argument by saying that 
``we want to find {\it the range} \ldots which contains the true value 
$\theta_\circ$ with probability $\beta$'', though rational 
people are at a loss in trying to convince themselves that the 
proposition ``the range contains $\theta_\circ$ with probability $\beta$''
does not imply ``$\theta_\circ$ is in that range with 
probability $\beta$''.}
about true values\cite{maxent98}), 
they fail in  ``difficult cases: 
small or unobserved signal, background larger than signal,
background not well known, and measurements near a physical 
boundary''\cite{CLW}. 

It is interesting to note that many of the above-cited papers 
on limits have been written in the wake of an article\cite{FC} which 
was promptly adopted by the
PDG\cite{PDG} as the longed for ultimate solution to
the problem, which could finally ``remove an original motivation
for the description of Bayesian intervals by the PDG''\cite{FC}.
However, although Ref.~\cite{FC}, thanks to the authority of the PDG,
has been widely used by many experimental teams to publish 
limits, even by people who did not understand the method
or were sceptical about it,\footnote{This 
non-scientific practice has been well expressed
by a colleague: ``At least we have a rule, no matter
if good or bad, to which we can adhere. Some of the limits 
have changed? You know, it is like when governments change 
the rules of social games: some win, some lose.'' 
When people ask me why I disagree with Ref.~\cite{FC}, 
I just encourage them to read  the paper carefully,
instead of simply picking a number from a table.}
that article has triggered a debate between those who simply 
object to the approach 
(e.g. Ref.~\cite{Zech}), those 
who propose other prescriptions (many of these authors do it with the 
explicit purpose of ``avoiding Bayesian contaminations''\cite{Punzi} 
or of ``giving a strong contribution to rid physics of Bayesian 
intrusions''\footnote{See 
Ref.\cite{Science} to get an idea of the present `Bayesian intrusion'
in the sciences, especially in those disciplines in which frequentistic
methods arose.}~\cite{Ciampolillo}), and those who just propose 
to change radically the path~\cite{ci,noi}. 

The present contribution to the debate, based on 
Refs.~\cite{ci,noi,maxent98,Higgs,YR,ajp}, 
is in the framework 
of what has been initially the physicists' 
approach to probability,\footnote{Insightful historical remarks 
about the correlation physicists-`Bayesians' (in the modern sense)
can be found in the first two sections of Chapter 10 of  
Jaynes' book~\cite{Jaynes}. For a more extensive account of  
the original approach of Laplace, Gauss and other physicists
and mathematicians, see Ref.~\cite{Hald}.} 
and which I maintain\cite{maxent98} 
is still the intuitive reasoning of the large majority
of physicists, despite  the `frequentistic intrusion'
in the form of standard statistical courses in the 
physics curriculum.
I will show by examples that 
an aseptic prior-free
assessment of `confidence' is a contradiction in terms
and, consequently, that {\it the} solution to the problem of 
assessing `objective' confidence limits does not exist. Finally,
I will show how it is possible,
nevertheless, to present search results 
in an  objective (in the sense this committing word is 
commonly perceived) and optimal way 
which satisfies the desiderata expressed in Section
\ref{sec:desiderata} section.
The price to pay is to remove the expression `confidence limit'
from our parlance and talk, instead, of `sensitivity bound' 
to mean a prior-free result. Instead, the expression
`probabilistic bound' should be used to assess how much 
we are really \underline{confident}, i.e. how much we believe,
that the quantity of interest is above or below the bound,
under clearly stated prior assumptions. 

The present paper focuses mostly on the `difficult cases'\cite{CLW}, 
which will be  classified as `frontier measurements'~\cite{priors},
characterized by an `open likelihood', as  will be better
specified in Section \ref{sec:routine},  where this situation 
will be compared to the easier case of `close likelihood'.
It will be shown why
 there are good reasons to present routinely 
the experimental outcome 
in two different ways for the two cases.

\section{DESIDERATA FOR AN OPTIMAL PRESENTATION OF SEARCH RESULTS}
\label{sec:desiderata}

Let us specify an optimal presentation
of a search result in terms of some desired properties. 
\begin{itemize}
\item
The way of reporting the result should not depend on 
whether the experimental team is more or less convinced 
to have found the signal looked for. 
\item
The report should allow an easy, consistent and efficient 
combination of all pieces of 
information which could come from several experiments,
search channels and running periods. By efficient I mean the following:
if many independent data sets each provide a little evidence in
favour of the searched-for signal, 
the combination of all data should enhance that hypothesis;
if, instead, the indications provided by the different
data are incoherent, their combination should result in 
stronger constraints on the intensity of the 
postulated process (a higher mass, a lower coupling,  etc.). 
\item
Even results coming from low sensitivity (and/or very noisy) 
data sets could be included in the combination, 
without them spoiling the quality of the result 
obtainable by the clean and high-sensitivity data sets alone.
If the poor-quality data carry the slightest piece of 
evidence, this information should play the correct role
of slightly increasing the global evidence.
\item
The presentation of the result 
(and its meaning) should not depend on the 
particular application (Higgs search, scale of contact-interaction,
proton decay, etc.).
\item
The result should be stated in such a way that it 
cannot be misleading. This requires that it should  easily
map into the natural categories developed by 
the human mind for uncertain events.   
\item
Uncertainties due to systematic effects of  
uncertain size should be included in a
consistent and (at least conceptually) simple way.
\item
Subjective contributions of the persons who provide the results 
should be kept at a minimum. These contributions
 cannot vanish, in the
sense that we have always to rely on the ``understanding,
critical analysis and integrity''~\cite{ISO} of the experimenters 
but at least the dependence on the believed values
of the quantity should be minimal.
\item
The result should summarize completely the experiment, and no
extra pieces of information (luminosity, cross-sections,
efficiencies, expected
number of background events, observed number of events)
should be required for further analyses.\footnote{For example, 
during the work for Ref.~\cite{Higgs}, we were
unable to use only the `results', and had to restart 
the analysis from the detailed pieces of information,
which are not always as detailed as one would need.
For this reason
we were quite embarrassed when, finally,  
we were unable to use consistently the 
information published by
one of the four LEP experiments.}
\item
The result should be ready to be turned into probabilistic 
statements, needed to form one's opinion about the 
quantity of interest or to take decisions.
\item
The result should not lead to paradoxical conclusions.  
\end{itemize}
 
\section{ASSESSING THE DEGREE OF CONFIDENCE}

As Barlow says~\cite{Barlow}, ``Most statistics 
courses gloss over the
definition of what is meant by {\it probability}, with at 
best a short mumble to the effect that there is no universal 
agreement. The implication is that such details are 
irrelevancies of concern only to long-haired philosophers, and need
not trouble us hard-headed scientists. This is short-sighted; 
uncertainty about what we really mean when we calculate 
probabilities leads to confusion and bodging, particularly 
on the subject of {\it confidence levels}. \ldots Sloppy thinking
and confused arguments in this area arise mainly from changing one's
definition of `probability' in midstream, or, indeed, of not 
defining it clearly at all.'' Ask your colleagues how they perceive 
the statement ``95\% confidence level lower bound of 77.5
GeV/$c^2$ is obtained for the mass of the Standard Model 
Higgs boson''\cite{LEP}. I conducted an extensive poll in 
July 1998, personally and by electronic mail. 
The result\cite{maxent98} is that for the large 
majority of people the above statement means that ``assuming the Higgs 
boson exists, we are 
95\% confident that the Higgs mass is above that limit, 
\underline{i.e.} the Higgs boson has 95\% chance (or probability)
of being on the upper side, and 5\% chance of being on the 
lower side,''\footnote{Actually, there were those who
refused to answer the question because ``it is going to be difficult
to answer'', and those who insisted on repeating the frequentistic
lesson on lower limits, but without being able to provide 
a convincing statement understandable to a scientific journalist
or to a government authority -- these were the terms of the question -- 
about the degree of confidence that the Higgs is heavier than the 
stated limit. I would like to report the 
latest reply to the poll, which arrived just the day
before this workshop: ``I apologize I never got around to answering your
mail, which I suppose you can rightly regard as 
evidence that the classical procedures are not trivial!''}, 
which is not what the operational definition
of that limit implies~\cite{LEP}.
Therefore, following the suggestion of Barlow~\cite{Barlow}, let us 
``take a look at what we mean by the term `probability' (and confidence)
before discussing the serious business of confidence levels.''
I will do this with some examples, referring to 
Refs.\cite{YR,ajp} for more extensive discussions and further examples. 

\subsection{Variations over a problem to Newton}

It seems\footnote{My source of information is Ref.~\cite{Glymour}.
It seems that Newton gave the `correct answer' - 
indeed, in this stereotyped problem there is {\it the}
correct answer.} 
that Isaac Newton was asked to solve the following problem. 
A man condemned to death has an opportunity of having 
his life saved and to be freed, depending on the outcome of an
uncertain event. The man can choose between three options: 
a) roll 6 dice, and be free if he gets `6' with one and only one 
die ($A$);
b) roll 12 dice, and be freed if he gets `6' with exactly 
2 dice; c) roll 18 dice, and be freed if he gets `6' in exactly 3 dice. 
Clearly, he will choose the event about which he is {\it more confident}
(we could also say the event which he considers {\it more probable}; 
the event {\it most likely to happen}; 
the event which {\it he believes mostly}; and so on). 
Most likely the condemned man is not able to solve the problem, 
but he certainly will understand Newton's suggestion 
to choose $A$, which gives him the {\it highest chance} to survive. 
He will also understand the statement that $A$ is about six times 
more likely than $B$ and thirty times more likely than $C$.
The condemned would perhaps ask Newton to give him some 
idea how likely the event $A$ is. A good answer would 
be to make a
comparison with a box containing 1000 balls, 94 of which are white. 
He should be so confident of surviving as of extracting a white ball
from the box;\footnote{The reason why any person is able 
to claim to be more confident of extracting a white ball
from the box that contains the largest fraction of white balls, 
while for the evaluation of the above events one has to `ask Newton',
does not imply a different perception of the `probability' in the two 
classes of events. It is only because the 
events $A$, $B$ and $C$ are complex events, the probability 
of which is evaluated from the probability of the elementary events
(and everybody can figure out what it means that the six faces of a die 
are equally likely) plus some combinatorics, for which 
some mathematical education is needed.} 
i.e. 9.4\% confident of being freed and 90.6\% 
confident of dying: not really an enviable situation, but better
than choosing $C$, corresponding to only 3 white balls in the box.

Coming back to the Higgs limit, are we
really honestly 95\% confident that the value of its mass 
is above the limit as  we are confident that a neutralino
mass is above its 95\% C.L. limit, as a given branching ratio is 
below its 95\% C.L. limit, etc., as we are confident 
 of extracting a white ball from a box which contains 
95 white  and 5 black balls? 

\begin{figure}
\begin{center}
\epsfig{file=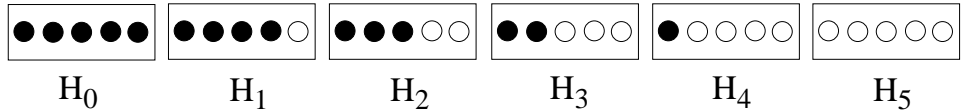,width=0.8\linewidth,clip=}
\end{center}
\vspace{-0.5cm}
\caption{\small A box has with certainty one of these six 
black and white ball compositions.
The content of the box is 
inferred by extracting at random a ball from the 
box then returning it to the box.
How confident are you initially 
 of each composition? How does your confidence change
after the observation of 1, 5 and 8 consecutive extractions of a 
black ball?}
\label{fig:six_compositions}
\end{figure}

Let us imagine now a more complicated situation, in which 
you have to make the choice (imagine for a moment
\underline{you} are the prisoner, just to be emotionally more
involved in this academic 
exercise\footnote{Bruno de Finetti used to say
that either probability concerns real events in which we
are interested, or it is nothing
\cite{deFinetti}.\label{fn:deFinetti}}).
A box contains with certainty 5 balls, with a white ball
content ranging  from 0 to 5, the remaining balls being 
black (see Fig.~\ref{fig:six_compositions}, and Ref.~\cite{ajp} 
for further variations on the problem.).
One ball is extracted at random, shown to you, and 
then returned to the box. The ball is {\bf black}.  
You get freed if you guess correctly the composition of the box. 
Moreover you are allowed to ask a question, to which the judges
will reply correctly if the question is pertinent 
and such that their answer does not indicate 
with certainty the exact content of the box. 

Having observed a black ball, 
the only certainty is that $H_5$ is ruled out. 
As far as  the other five possibilities are concerned, 
a first idea would be to be more confident about the 
box composition which has more black balls ($H_0$), since 
this composition gives the highest chance of extracting this  
colour. Following this reasoning, 
the confidence in the various box compositions would be 
proportional to  their black ball content. But it is not 
difficult to understand that this solution is obtained by 
assuming that the compositions are considered a priori 
equally possible. However, this condition was not stated explicitly in 
the formulation of the problem. How was the box prepared? 
You might think of an initial situation of 
six boxes each having a different composition. 
But you might  also think that the balls were picked at random from
 a large bag containing a roughly equal proportion 
of white and black balls.
Clearly, the initial 
situation changes. In the second case the composition $H_0$ is initially 
so unlikely that, even after having extracted a black ball,
it remains not very credible. As  eloquently
said by Poincar\'e~\cite{Poincare}, ``an effect may be
produced by the cause $a$ or by the cause $b$. The effect 
has just been observed. We ask the probability that it is 
due to the cause $a$. This is an {\it a posteriori}
probability of cause. But I could not calculate it, if
a convention more or less justified did not tell me
in advance what is the {\it priori} probability
for the cause $a$ to come into play. I mean the probability
of this event to some one who had not observed the effect.'' 
\begin{figure}
\begin{center}
\begin{tabular}{|c|c|} \hline
\multicolumn{2}{|c|}
{\epsfig{file=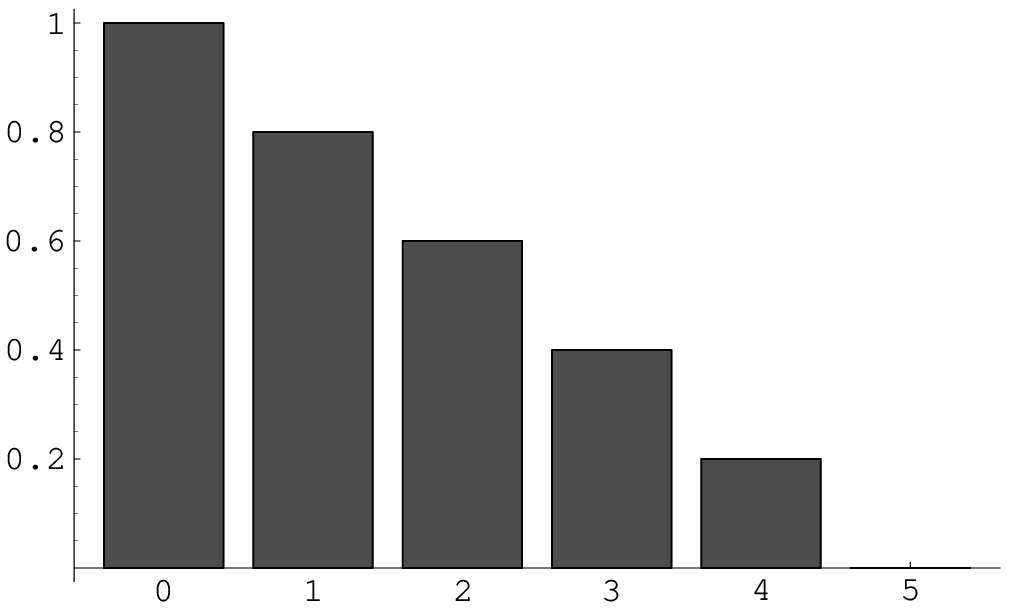,width=0.40\linewidth,clip=}}
\\ \hline
\epsfig{file=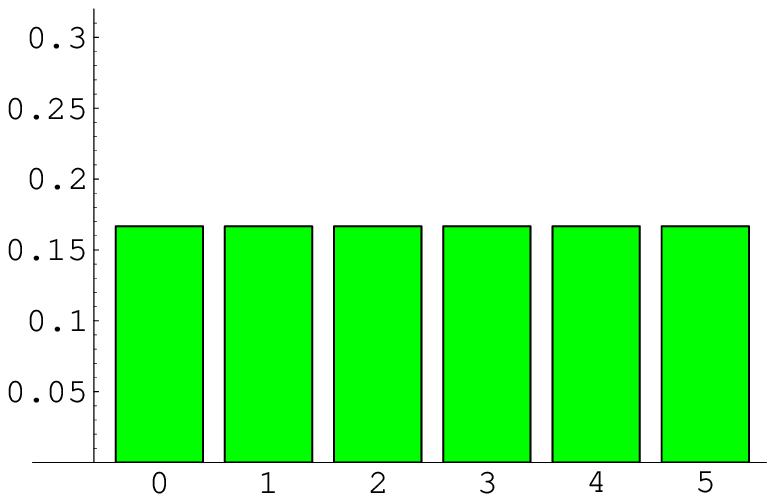,width=0.40\linewidth,clip=} &
\epsfig{file=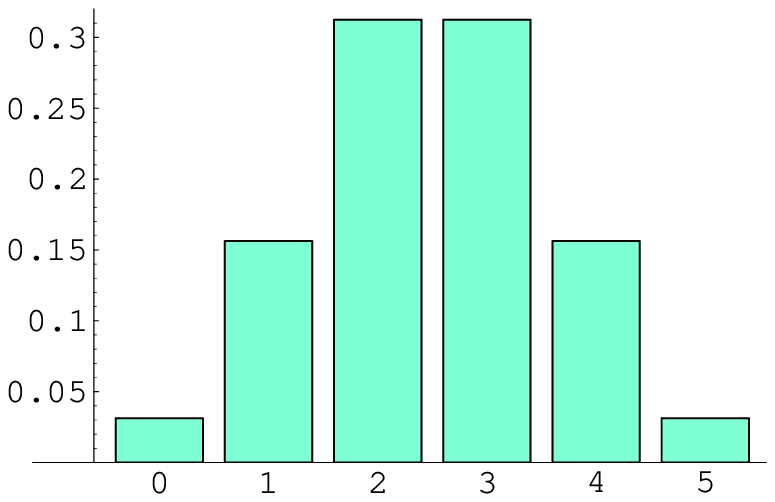,width=0.40\linewidth,clip=} \\ \hline
\epsfig{file=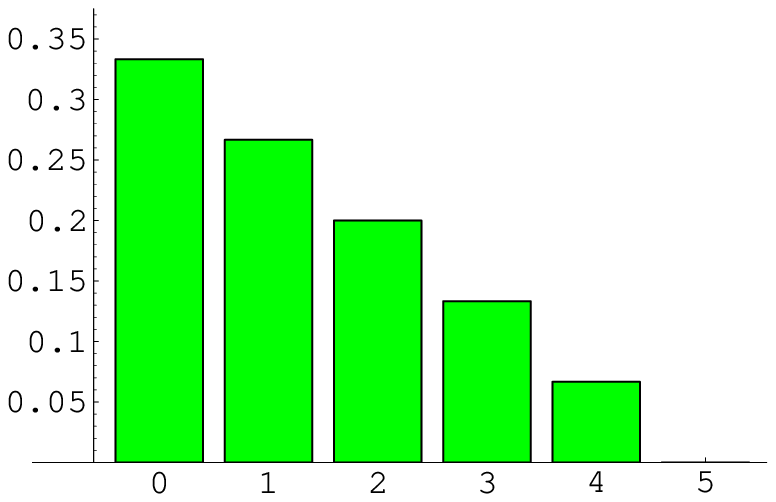,width=0.40\linewidth,clip=} &
\epsfig{file=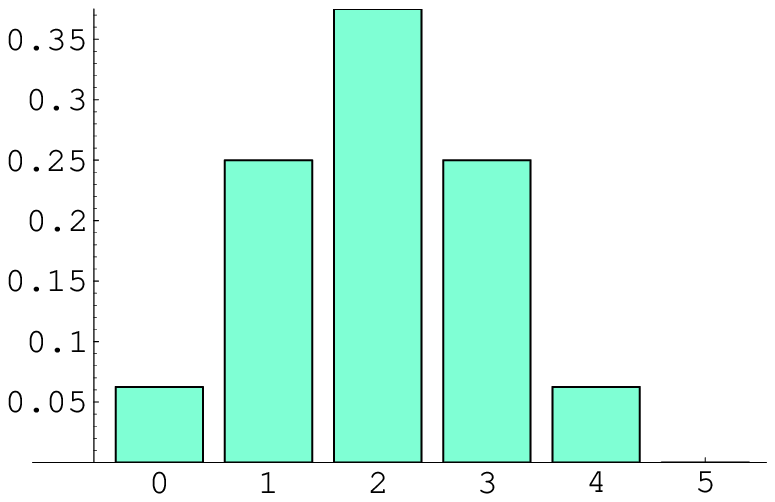,width=0.40\linewidth,clip=} \\ \hline
\epsfig{file=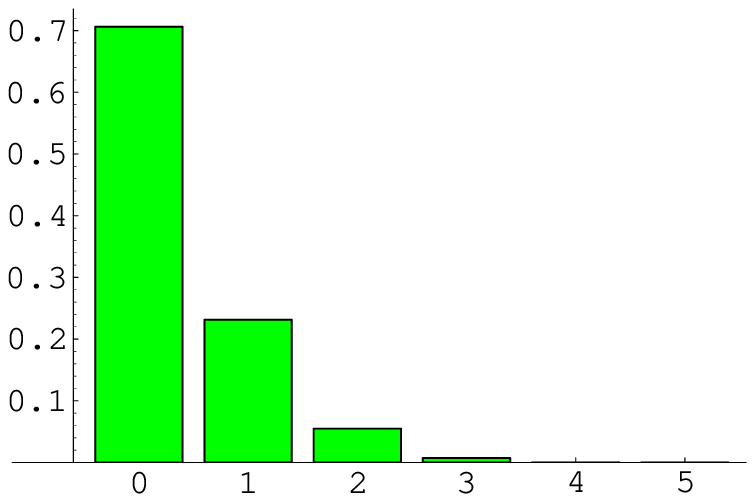,width=0.40\linewidth,clip=} &
\epsfig{file=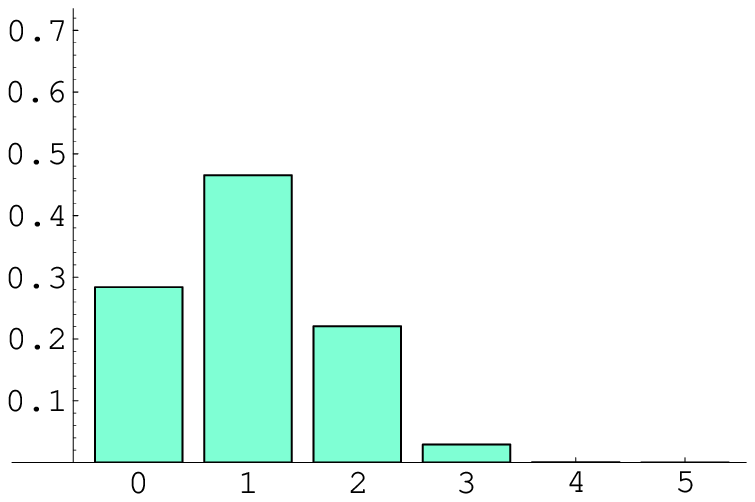,width=0.40\linewidth,clip=} \\ \hline
\epsfig{file=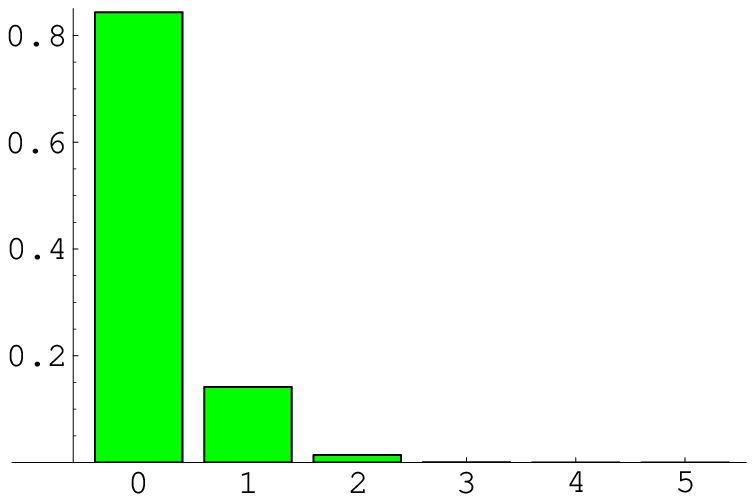,width=0.40\linewidth,clip=} &
\epsfig{file=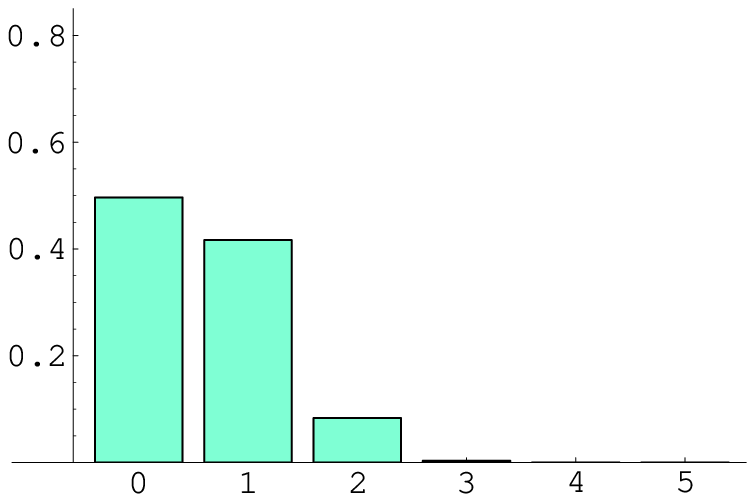,width=0.40\linewidth,clip=} \\ \hline
\end{tabular}
\end{center}
\caption{\small Confidence in the box contents as a function of
prior and observation (see text).} 
\vspace{-22.4cm}\hspace{+2.3cm}$P(\mbox{Black}\,|\,H_i)$

\vspace{+1.7cm}\hspace{+11.3cm}$H_i$

\vspace{+1.3cm}\hspace{+2.5cm}$P(H_i)$\hspace{+2.4cm}prior 1
\hspace{+2.3cm}$P(H_i)$ \hspace{+2.7cm}prior 2

\hspace{+5.7cm}(uniform)\hspace{+5.7cm}(binomial)

\vspace{+3.5cm}\hspace{+5.9cm}1 Black
\hspace{+5.9cm}1 Black

\vspace{+4.0cm}\hspace{+5.9cm}5 Black
\hspace{+5.9cm}5 Black

\vspace{+4.0cm}\hspace{+5.9cm}8 Black
\hspace{+5.9cm}8 Black

\vspace{+2.1cm}\hspace{+7.1cm}$H_i$ \hspace{+6.3cm}$H_i$

\vspace{+2.2cm}
\label{fig:all_black}
\end{figure}
The observation alone is not enough to state how much 
one is confident about something. 

The  proper way 
to evaluate the level of confidence,  which 
takes into account (with the correct weighting)
experimental evidence and prior knowledge, is 
recognized to be Bayes' theorem:\footnote{See Ref.~\cite{ajp}
for a derivation of Bayes' theorem based on the 
box problem we are dealing with.}
\begin{equation}
P(H_i\,|\,E) \propto P(E\,|\,H_i)\cdot P_\circ(H_i)\,,
\label{eq:Bayes}
\end{equation}
where $E$ is the observed event (black or white), 
$P_\circ(H_i)$ is the initial
(or a priori) probability of $H_i$ (called often simply `prior'), 
$P(H_i\,|\,E)$ is the final (or `posterior') probability, 
and $ P(E\,|\,H_i)$ is the `likelihood'. 
The upper plot of Fig.~\ref{fig:all_black} shows the 
likelihood $P(\mbox{Black}\,|\,H_i)$
of observing a black ball assuming each possible composition. 
The second pair of plots shows the two priors considered 
in our problem. The final probabilities are 
shown next. We see that the two solutions are quite different, 
as a consequence of different priors. So a good question 
to ask the judges would be how the box was prepared.
 If they say it was uniform, bet your life 
on $H_0$. If they say the five balls were extracted from 
a large bag, bet on $H_2$. 

Perhaps the judges might be so clement as  
to repeat the extraction (and subsequent reintroduction) 
several times. Figure \ref{fig:all_black} shows what happens 
if five or height consecutive black balls are observed. The evaluation
is performed by iterating Eq.~(\ref{eq:Bayes}):
\begin{equation}
P_n(H_i\,|\,E) \propto P(E_n\,|\,H_i)\cdot P_{n-1}(H_i)\,.
\label{eq:Bayes_iter}
\end{equation}
If you are 
convinced\footnote{And if you have doubts about the 
preparation? The probability rules teach us what to do. 
Calling $U$ (uniform) and $B$ (binomial) the two preparation 
procedures, with probabilities $P(U)$ 
and $P(B)$, we have
$P(H\,|\,\mbox{obs}) = P(H\,|\,\mbox{obs},U)\cdot P(U) +
                        P(H\,|\,\mbox{obs},B)\cdot P(B)\,.$}
 that the preparation procedure is the binomial one
(large bag), you still consider $H_1$ more likely than $H_0$,
even after five consecutive observations. 
Only after eight consecutive extractions of a black ball  are you mostly
confident about $H_0$ independently of how much you believe 
in the two preparation procedures (but, obviously, you might imagine --
and perhaps even believe in -- more fancy 
preparation procedures which still 
give different results). 
%
%
%
%
After many extractions we are practically sure
of the box content, as we shall see in a while, 
though we can never be certain. 

Coming back to the limits, imagine now an experiment 
operated for a very short time at LEP200 and reporting no 
four-jet events, no deuterons, no zirconium and 
no Higgs candidates (and you might add something even more fancy, 
like events with 100 equally energetic photons, or some 
organic molecule). How could the 95\% upper limit to the
rate of these events be the same? What does it mean 
that the 95\% upper limit calculated automatically
should give us the same confidence
for all rates, independently of what the events  are? 

\subsection{Confidence versus evidence}

The fact that the same (in a crude statistical sense)
observation does not lead to the same 
assessment of confidence is rather well understood
by physicists:
 a few pairs of photons
clustering in invariant mass around 135 MeV have a high
chance of coming from a  $\pi^\circ$; more events clustering 
below 100 MeV are certainly background (let us consider 
a well calibrated detector); a peak in  invariant 
mass in a new energy domain might be seen as a hint of new physics, 
and distinguished theorists consider it worth serious
speculation.
 The difference between the three cases is the prior
knowledge (or scientific prejudice). Very often we share 
more or less the same prejudices, and consequently we will all agree 
on the conclusions. But this situation is rare
in frontier science, and the same observation does not produce in all 
researchers the same confidence.
A peak can be taken more or less seriously 
depending on whether it is expected, it fits well in the overall
theoretical picture, and does not contradict other observations.
Therefore it is important to try to separate experimental 
evidence from the assessments of confidence. This 
separation is done in a clear
and unambiguous way in the Bayesian approach. Let us illustrate
it by continuing with the box example. 
Take again Eq.~(\ref{eq:Bayes}). Considering any two hypotheses
$H_i$ and $H_j$, we have the following relation between prior and
posterior {\it betting odds}:   
\begin{equation}
\frac{P(H_i\,|\,E)}{P(H_j\,|\,E)} =
\underbrace{\frac{P(E\,|\,H_i)}{P(E\,|\,H_j)}}_{\mbox{\it Bayes factor}}
\!\!\cdot\, \frac{P_\circ(H_i)}{P_\circ(H_j)}\,.
\label{eq:Bayes_factor}
\end{equation}
This way of rewriting the Bayes's theorem shows 
how the final odds can be factorized into 
 prior odds and experimental evidence, the latter expressed 
in terms of the so-called Bayes factor~\cite{Lavine}.
The 15 odds 
of our example 
are not independent, and can be expressed with respect to
a reference box composition which has a non-null likelihood. 
The natural choice to analyse the problem of consecutive 
black ball extractions is
\begin{equation}
{\cal R}(H_i\,;\,\mbox{Black}) = \frac{P(\mbox{Black}\,|\,H_i)}
                                  {P(\mbox{Black}\,|\,H_0)}\,,
\label{eq:R_box}
\end{equation}
which is, in this particular case, 
numerically identical to $P(\mbox{Black}\,|\,H_i)$, 
since $P(\mbox{Black}\,|\,H_0)=1$, and then it can be read from the 
top plot of Fig.~\ref{fig:all_black}. The function ${\cal R}$ 
can be seen as a `relative belief updating ratio'\cite{noi}, 
in the sense that it tells us how the beliefs \underline{must} 
be changed after the observation, though it cannot determine 
univocally their values. Note that the way the update is done is, instead,
univocal and not subjective, in the sense that Bayes' theorem 
is based on logic, and rational people cannot disagree. 
It is also obvious what happens when many consecutive back 
balls are observed. The iterative application of Bayes' theorem
[Eq.~(\ref{eq:Bayes_iter})] leads to the following overall ${\cal R}$:
\begin{equation}
{\cal R}(H_i\,;\,\mbox{Black},n) = \left[\frac{P(\mbox{Black}\,|\,H_i)}
                                              {P(\mbox{Black}\,|\,H_0)}
                                   \right]^n\,.
\label{eq:R_box_n}
\end{equation}
For large $n$ all the odds with respect to $H_0$ go to zero,
i.e. $P(H_0\rightarrow 0$\,.

We have now our logical and mathematical apparatus ready. 
But before moving to the problem of interest, 
let us make some remarks 
on terminology, on the meaning of subject probability, 
and on its interplay with odds in betting and expected frequencies. 

\subsection{Confidence, betting odds and expected frequencies}
\label{ss:confidence} 

I have used on purpose several words and expressions
 to mean essentially the same thing:
likely, probable, credible, (more or less) possible, 
plausible, believable, and their associated nouns; to be more or less
confident about, to believe more or less, to trust more or
less, something, and their associated
nouns; to prefer to bet on an outcome rather than another one, 
 to assess betting odds, and so on. I could also use expressions 
involving expected frequencies of outcomes of apparently similar
situations. The perception of probability would remain the same, 
and there would be no ambiguities or paradoxical conclusions. 
I refer to Ref.\cite{ajp} for a more extended, though still 
concise, discussion on the terms. 
I would like only to sketch here 
some of the main points, as a summary of the previous sections.
\begin{itemize}
\item
The so-called subjective probability is based on the 
acknowledgement that the concept of probability is primitive,
i.e. it is meant as the degree of belief developed by the human mind
in a condition of uncertainty, no matter what we call it
(confidence, belief, probability, etc) or how we evaluated it
(symmetry arguments, past frequencies, Bayes' theorem, 
quantum mechanics formulae\cite{clw_qm}, etc.). Some
argue that the use of beliefs is not scientific. 
I believe, on the other hand, that
{\it ``it is scientific only to say what is more likely and what it is 
less likely''}~\cite{Feynman}.  
\item
The odds in an `coherent bet' (a bet such that the person 
who assesses its odds has no preference in either direction)
can be seen as the normative 
rule to force people to assess honestly their degrees of belief
`in the most objective way' 
(as this expression is usually perceived). 
This is the way that Laplace used to report his result 
about the mass of Saturn: ``it is a bet of 10,000 to
1 that the error of this result is not 1/100th of its values'' 
(quote reported in Ref.~\cite{Sivia}). 
\item
Probability statements have to satisfy the basic rules of 
probability, usually known as axioms. Indeed, the basic rules
can be derived, as theorems, from the operative 
definition of probability through a coherent bet.
The probability rules, based on the axioms and on logic's
 rules, allows the probability 
assessments to be propagated to logically connected events. 
 For example, if one claims to be $xx\%$ confident about $E$, 
one should feel also $(100-xx)\%$ confident about $\overline{E}$.   
\item
The simple, stereotyped cases   
of regular dice and urns of known composition
can be considered as calibration tools to
assess the probability, 
in the sense that all rational people will agree.
\item
The probability rules, and in particular Bernoulli's theorem,
relate degrees of belief to expected frequencies, if 
we imagine repeating 
the experiment many times under exactly the same conditions
of uncertainty (not necessarily under the same physical conditions). 
\item
Finally, Bayes' theorem is the logical tool to update the beliefs
in the light of new information. 
\end{itemize}
As an example, let us imagine
the  event $E$, which is considered  95\% probable 
(and, necessarily, the  opposite event $\overline{E}$ is 5\% probable). 
This belief can be expressed in many different ways,  all 
containing the same degree of  uncertainty  : 
\begin{itemize}
\item
I am 95\% confident about $E$ and 5\% confident about $\overline{E}$.
\item
Given a box containing 95 white and 5 black balls, 
I am as  confident that $E$ will happen, as that the colour 
of the ball will be white.
I am as confident about $\overline{E}$ 
 as of extracting a black ball. 
\item
I am ready to place a 19:1 bet\footnote{See Ref.\cite{ajp} for comments 
on decision problems involving subjectively-relevant 
amounts of money.} 
on $E$, or a 1:19 on $\overline{E}$. 
\item
Considering a large number $n$ of events $E_i$, even related 
to different phenomenology and each having 95\% probability,
I am highly confident\footnote{It is in my opinion 
very important to understand the 
distinction between the use of this frequency-based 
expression of probability and frequentistic approach 
(see comments in Refs. \cite{ajp} and \cite{YR}) or 
frequentistic coverage (see Section 8.6 of Ref.~\cite{YR}).
I am pretty sure that most physicists who declare to be frequentist
do so on the basis of educational conditioning and because they
are accustomed to assessing beliefs (scientific opinion, or whatever)
in terms of expected frequencies. The crucial point which makes 
the distinction is it to ask oneself if it 
is sensible to speak about probability of
true values, probability of theories, and so on. 
There is also a class of sophisticated people who think there
are several probabilities. For comments on this latter attitude,
see Section 8.1 of Ref.~\cite{YR}.}  
that the relative frequency of the events   
which will happen will be very close to 95\%
(the exact assessment of my confidence can be 
evaluated using the binomial distribution). 
If $n$ is very large, 
I am practically sure that the relative frequency will be
equal to 95\%, but I am never certain, 
unless $n$ is `infinite', but this is no longer a real problem,
in the sense of the comment in 
footnote \ref{fn:deFinetti}
(``In the long run we are all dead''~\cite{Keynes}). 
\end{itemize}
Is this  how our confidence limits 
from particle searches are perceived? 
Are we really 5\% confident that the quantity of interest is on the
5\% side of the limit? Isn't it strange that out of the several 
thousand limits from searches published in recent decades  
nothing  has ever shown up on the 5\% side? 
In my opinion, the most embarrassing situation comes from 
the Higgs boson sector. A 95\% C.L. upper limit is obtained 
from radiative corrections, while a 95\% C.L. limit  
comes from direct search. Both results are presented with the 
same expressions, only `upper' being replaced by `lower'. 
But their interpretation is completely different. In the first case
it is easy to show~\cite{Higgs_clw} that, using the almost parabolic  
result of the $\chi^2$ fit in $\ln(M_H)$ and uniform prior in 
$\ln(M_H)$, we can really talk about `95\% confidence that 
the mass is below the limit', or that `the Higgs mass has equal
chance of being on either side of the value of minimum $\chi^2$',
and so on, in the sense described in this section. 
This is not true in the second 
case.  Who is really 5\% confident that the mass is below the limit?   
How can we be 95\% confident that the mass is above the limit
without an upper bound? Non-misleading levels 
of confidence on the statement $M_H>M_\circ$ can be assessed only  
by using the information coming from
precision measurement, which rules out very large (and also very small)
values of the Higgs mass (see
Refs.~\cite{Erler,Higgs,Higgs_clw}. 
For example, when we say \cite{Higgs_clw} that 
the median of the Higgs mass p.d.f. is 150\,GeV, we mean that, 
to best of our knowledge, we regard the two events $M_H<150$ and 
$M_H>150$ as equally likely, like the two faces of a regular coin. 
Following Laplace, we could state our confidence claiming 
that `is a bet of 1 to 1  that $M_H$ is below 150 GeV'.    

\section{INFERRING THE INTENSITY OF POISSON PROCESSES AT THE LIMIT
OF THE DETECTOR SENSITIVITY AND IN THE PRESENCE OF BACKGROUND}

As a master example of frontier measurement, let us take the 
same  case study as in Ref.~\cite{noi}. We shall focus then
on the inference of the rate of gravitational wave (g.w.)
bursts measured by coincidence analysis of g.w. antennae. 

\subsection{Modelling the inferential process}

Moving from the box example to 
the more interesting physics
case of g.w. burst is quite straightforward. 
The six hypotheses $H_i$, playing the role of causes,  
are  now replaced by the infinite values of the rate $r$.
The two possible outcomes 
 black and white now 
become  the  number of candidate events
($n_c$).  There is also  an extra 
ingredient which comes into play:
a candidate event could  come  from background rather than
from g.w.'s 
(like a black ball that could be extracted by a judge-conjurer
from his pocket rather than from the box\ldots). 
Clearly, if we understand well  the experimental apparatus, we 
must have 
some idea of the background rate $r_b$. Otherwise, it is better to 
study further the performances of the detector, before trying to 
infer anything. Anyhow, unavoidable residual uncertainty 
on $r_b$ can 
be handled consistently (see later). 
Let us summarize our ingredients in terms 
of Bayesian inference.  
\begin{itemize}
\item
The physical quantity of interest, and with respect to which 
we are in the  state of greatest uncertainty, is the  
g.w. burst rate $r$.
\item
We are rather sure about the expected rate of background 
events $r_b$ (but not about the number of events due to background 
which will actually be observed).
\item
What is certain\footnote{Obviously the problem can be 
complicated at will, considering for example that $n_c$
was communicated to us in a way, or by somebody, which/who is  
not 100\% reliable. A probabilistic theory can include
this possibility, but this goes beyond the purpose
of this paper. See e.g. Ref.~\cite{BN} 
for further information on probabilistic networks.} 
is the number $n_c$ 
of coincidences which have been observed.
\item
For a given hypothesis $r$  
the number of coincidence events which can 
be observed in the observation time $T$ is described by a Poisson 
process having an intensity which is the sum
of that due to background and that due to signal. Therefore the likelihood
is
\begin{equation}
P(n_c\,|\,r,r_b) = f(n_c\,|\,r,r_b) =
                  \frac{e^{-(r+r_b)\,T}((r+r_b)\,T)^{n_c}}{n_c!}\,.
\label{eq:likr}
\end{equation}
\end{itemize}
Bayes' theorem applied to probability functions and 
probability density functions (we use the same symbol for both), written 
in terms of the uncertain quantities of interest, is
\begin{equation}
f(r\,|\,n_c,r_b) \propto f(n_c\,|\,r,r_b)\cdot f_\circ(r)\,.
\label{eq:Bayes1}
\end{equation}
At this point, it is now clear that if we want to assess 
our confidence we need to choose some prior. We shall come back to
this point later. Let us see first, following the box problem, how it
is possible to make a prior-free presentation of the result.

\subsection{Prior-free presentation of the experimental evidence}

Also in the continuous case we can factorize the 
prior odds and experimental evidence, 
and then arrive at an ${\cal R}$-function similar to
Eq.~(\ref{eq:R_box}): 
%
\begin{equation}
{\cal R}(r;n_c,r_b) = \frac{f(n_c\,|\,r,r_b)}{f(n_c\,|\,r=0,r_b)}\,.
\label{eq:rbur_def}
\end{equation}
The function ${\cal R}$ has nice 
intuitive interpretations which can be highlighted by 
rewriting the ${\cal R}$-function in the following way 
[see Eq.~(\ref{eq:Bayes1})]: 
\begin{equation}
{\cal R}(r;n_c,r_b) = \frac{f(n_c\,|\,r,r_b)}{f(n_c\,|\,r=0,r_b)} =
\frac{f(r\,|\,n_c,r_b)}{f_\circ(r)}\left/
\frac{f(r=0\,|\,n_c,r_b)}{f_\circ(r=0)}\right.\,.
\label{eq:rbur}
\end{equation} 
${\cal R}$ has the probabilistic interpretation of 
`relative belief
updating ratio', or the geometrical interpretation
of `shape distortion function' of the
probability density function. 
 ${\cal R}$ goes to 1 for $r\rightarrow 0$, i.e. in the asymptotic region
in which the experimental sensitivity is lost. As long as it is 1,
the shape of the p.d.f. (and therefore the relative probabilities in that 
region) remains unchanged.  In contrast,  in the limit 
${\cal R}\rightarrow 0$ (for large $r$) 
the final p.d.f. vanishes, i.e. the beliefs go 
to zero no matter how strong they were before. 
For the Poisson process we are considering, the relative 
${\cal R}$-function becomes
\begin{equation}
{\cal R}(r;n_c,r_b,T) = e^{-r\,T}\left(1+\frac{r}{r_b}\right)^{n_c}\,,
\label{eq:erre}
\end{equation}
with the condition $r_b>0$ if  $n_c >0$.
The case $r_b=n_c=0$ 
yields ${\cal R}(r) = e^{-r}$, obtainable starting directly 
from Eq. (\ref{eq:rbur_def}) and
Eq. (\ref{eq:likr}). Also the case 
$r_b\rightarrow \infty$ has to be evaluated directly from 
the definition of ${\cal R}$ and from the likelihood, 
yielding ${\cal R}=1\ \forall\, r$. Finally, 
the case $r_b=0$
and $n_c>0$ makes $r=0$ impossible, 
thus making the likelihood closed also on the 
left side (see Section \ref{sec:routine}).
In this case the discovery is certain, though the exact value 
of $r$ can be still rather uncertain. 
Note, finally, that if $n_c=0$ the ${\cal R}$-function does 
not depend on $r_b$, which might seem a bit surprising
at a first sight (I confess that have been puzzled
for years about this result which was formally correct,
though not intuitively obvious. 
Pia Astone has finally shown at this workshop 
that things must go logically this way~\cite{Pia}.)

\begin{figure}
\begin{center}
\epsfig{file=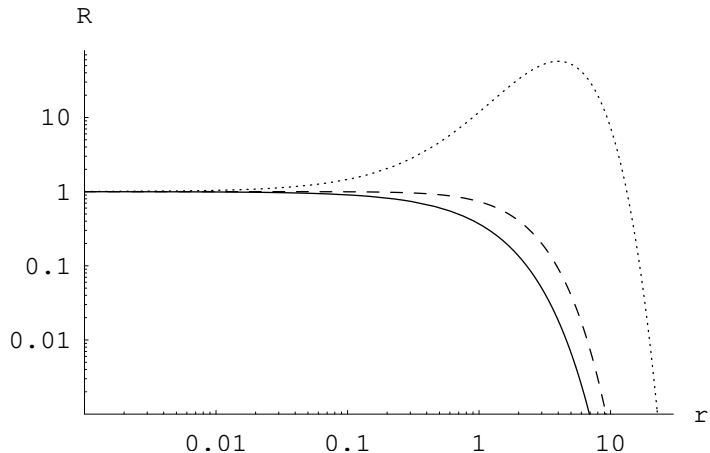,clip=,width=0.60\linewidth} 
\end{center}
\vspace{-0.4cm}
\caption{\small Relative belief updating ratio ${\cal R}$'s
for the Poisson intensity parameter $r$, in units of events
per month
evaluated from an expected rate of background events $r_b=1$ 
event/month
and the following numbers of observed events: 0 (continuous);
1 (dashed); 5 (dotted).}
\label{fig:rbur}
\end{figure}

A  numerical example will illustrate the nice features 
of the ${\cal R}$-function. 
Consider $T$ as unit time (e.g. one month), 
a background rate $r_b$ such that $r_b\times T=1$,
and the following hypothetical observations: 
 $n_c=0$;  $n_c=1$; $n_c=5$. The resulting ${\cal R}$-functions
are shown in Fig.~\ref{fig:rbur}.  The abscissa 
has been drawn in a log scale to make it clear that several orders 
of magnitude are involved. 
These curves transmit the result of the experiment 
immediately and intuitively.
Whatever one's beliefs on $r$ were before the data, these curves
show how one must change them.
The beliefs one had 
for rates far above 20 events/month are killed by the 
experimental result.
If one believed strongly that the rate had to be below 
0.1 events/month, the data are irrelevant.
The case in which no candidate events
have been observed gives the strongest constraint on the 
rate.
The case of five candidate events over an expected background of
one produces a  
 peak of ${\cal R}$ which  corroborates
 the beliefs 
around 4 events/month only if there were sizable 
prior beliefs in that region (the question 
of whether do g.w. bursts exist at all is discussed in Ref.~\cite{noi}).

Moreover there are some computational advantages in reporting 
the ${\cal R}$-function as a result of a search experiment:
The comparison between different 
results given by the ${\cal R}$-function can be perceived 
better than if these results were presented in terms of 
absolute likelihood.
Since ${\cal R}$ differs from the likelihood only 
by a factor, it can be used directly in  Bayes' theorem, 
which does not depend on constant factors, whenever 
probabilistic considerations are needed:
$
$
The combination of different independent 
results on the same
quantity $r$ 
can be done straightforwardly by multiplying  individual
${\cal R}$ functions;
note that a very noisy and/or low-sensitivity data set results in 
${\cal R}=1$ in the region where the good data sets yield 
an ${\cal R}$-value varying from 1 to 0, and then it does not
affect the result. 
One does not need to decide a priori if one wants to make a
`discovery' or an `upper limit' analysis: 
the ${\cal R}$-function
represents the most unbiased way of presenting the results
and everyone can draw their own conclusions.

Finally, uncertainty due systematic effects (expected background,
efficiency, cross-section, etc.)
can be taken 
into account in the likelihood using the laws of 
probability~\cite{noi} (see also Ref.~\cite{asym}).

\section{SOME EXAMPLES OF ${\cal R}$-FUNCTION BASED ON REAL DATA}

\begin{figure}
\begin{center}
\hspace{0.6cm}\epsfig{file=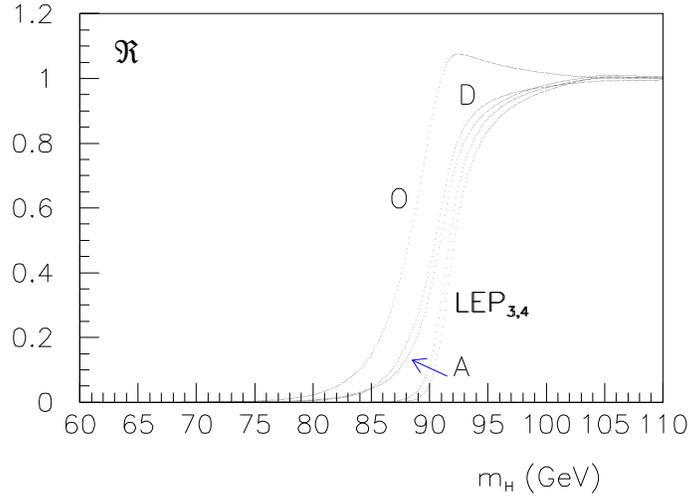,width=0.7\linewidth,clip=}
\end{center}
\vspace{-0.6cm}
\caption{\small ${\cal R}$-function reporting results on 
Higgs direct search from the
reanalysis of Ref.~\cite{Higgs}. A, D and O stand for 
ALEPH, DELPHI and OPAL. Their combined result is indicated 
by LEP$_3$. The full combination (LEP$_4$) 
was obtained by assuming for L3 a behaviour equal to the average 
of the others experiments.}
\label{fig:R_higgs}
\end{figure}

\begin{figure}
\begin{center}
\epsfig{file=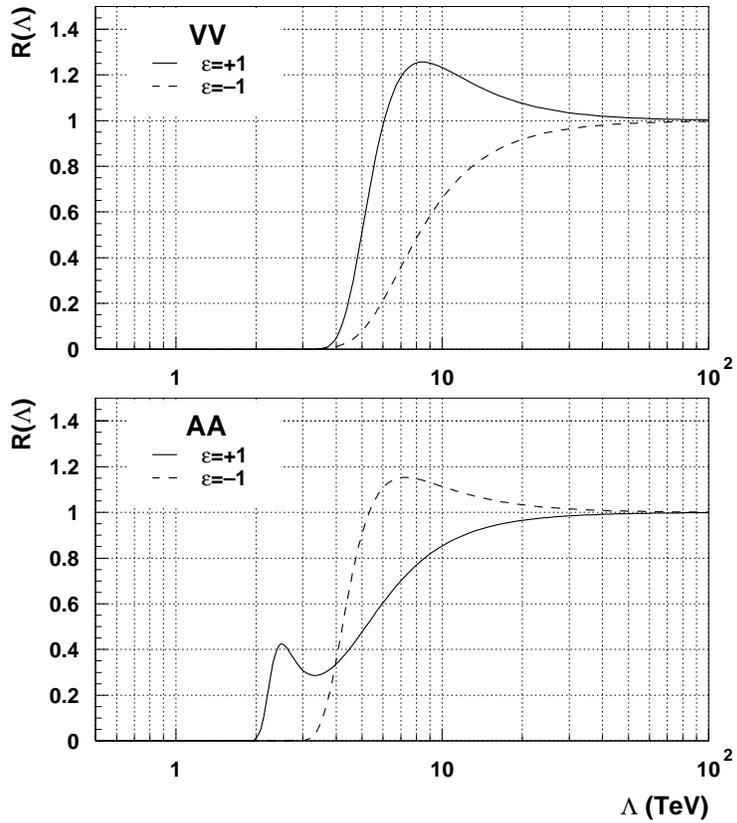,width=0.6\linewidth,clip=}
\end{center}
\vspace{-0.4cm}
\caption{\small ${\cal R}$-functions reporting results on search
for contact interactions~\cite{zeus}. The ZEUS paper 
contains the detailed information to obtain these curves,
as well as those relative to other couplings.}
\label{fig:R_zeus}
\end{figure}

The case study described till now is based on a toy model simulation. 
To see how the proposed method provides 
the experimental evidence in a clear way 
we show in Figs. \ref{fig:R_higgs} and \ref{fig:R_zeus}
${\cal R}$-functions based on real data. 
The first is a reanalysis
of Higgs search data at LEP\cite{Higgs}; the second comes 
from the search for contact interactions at HERA made 
by ZEUS\cite{zeus}. 
The extension of Eq.~(\ref{eq:rbur_def}) to the most general case
is 
\begin{equation}
{\cal R}(\mu;\mbox{data}) = 
\frac{f(\mbox{data}\,|\,\mu)}
{f(\mbox{data}\,|\,\mu_{\mbox{ins}})}\,,
\label{eq:rbur_gen}
\end{equation}
%
where $\mu_{\mbox{ins}}$ stands for the asymptotic insensitivity
value (0 or $\infty$, depending on the physics case)
of the generic quantity $\mu$. 
Figures \ref{fig:R_higgs} and \ref{fig:R_zeus} show clearly 
what is going on, namely which values are practically ruled out
and which ones are inaccessible to the experiment. 
The same is true for the result of a neutrino oscillation 
experiment reported two-dimensional ${\cal R}$-function~\cite{Doucet}
(see also Ref.~\cite{Eitel}).

\section{SENSITIVITY BOUND VERSUS PROBABILISTIC BOUND}

At this point, it is rather evident from Figs. \ref{fig:rbur},
\ref{fig:R_higgs} and \ref{fig:R_zeus} how we can 
summarize the result with a single number  which gives an idea 
of an upper or lower bound. 
In fact, 
although the ${\cal R}$-function 
represents the most complete 
and unbiased way of reporting the result, it might also
 be convenient to express with just one number
the result of a search which is considered 
by the researchers to be unfruitful.  
This number can be any value chosen by convention 
in the region where ${\cal R}$ has a transition
from 1 to 0. 
This value would then delimit (although roughly)   
 the region of the  
values of the quantity which are definitively excluded
from the region in which the experiment can say 
nothing. The meaning of this bound is not that 
of a probabilistic limit, but of a 
wall\footnote{In most cases it is not a sharp solid wall.
A hedge might be more realistic, and indeed more poetic: 
{\it ``Sempre caro mi fu quell'ermo colle, /
E questa siepe, che da tanta parte /
Dell'ultimo orizzonte il guardo esclude''} 
(Giacomo Leopardi, {\sl L'Infinito}). The exact position
of the hedge doesn't really matter, if we think that on the other
side of the hedge there are infinite orders of 
magnitude to which we are blind.}
which separates the region in which we are, and where 
we see nothing, from the the region we cannot 
see. We may take as the conventional position of the wall 
the point where ${\cal R}(r_L)$ equals $50\%$, $5\%$ or  $1\%$
of the insensitivity plateau.
What is important 
is not to call this value a bound at a given probability level
(or at a given confidence level -- the perception
of the result by the user will be the same!~\cite{maxent98}). 
A possible unambiguous name, corresponding to what this number
indeed is,  could be
`standard sensitivity bound'. 
As the conventional level, our suggestion is to choose
${\cal R} = 0.05$~\cite{noi}. 

Note that it does not make much sense to give the 
standard sensitivity bound with many significant digits.
The  reason becomes clear by observing 
Figs.~\ref{fig:rbur}--\ref{fig:R_zeus}, in particular 
Fig.~\ref{fig:R_zeus}. 
I don't think that there will be a single physicist
who, judging from the figure, 
believes that there is a substantial difference
concerning the scale of a postulated contact interaction
for  $\epsilon=+1$ and  $\epsilon=-1$\,. Similarly, 
looking at Fig.~\ref{fig:rbur}, the observation of 0 events, 
instead of 1 or 2, should not produce a significant  
modification of our opinion about g.w. burst rates.
What really matters is the order of magnitude of the bound
or, depending on the problem, the order of magnitude of 
the difference between the bound and the kinematic 
threshold (see discussion in Sections 9.1.4 and
 9.3.5 of Ref.~\cite{YR}). I have the impression 
that often the determination of a limit is considered
as important as the determination of the value of
a quantity. A limit should be considered on the same footing as 
an uncertainty, not as a true value. We can, at least in principle, 
improve our measurements and 
increase the accuracy on 
the true value. This reasoning cannot be applied to bounds.  
Sometimes I have the 
feeling that when some talk about a `95\% confidence limit',
they think as if they were `95\% confident \underline{about}
the limit'. It seems to me that for this reason
some are disappointed to see upper limits on the Higgs mass
fluctuating, in contrast to lower limits which are 
more stable and in constant
increase with the increasing available energy. In fact,
as said above, these two 95\% C.L. limits don't have the same meaning.
It is quite well understood by experts that lower 95\% C.L. limits
are in practice $\approx 100\%$ probability limits,
and they are used in theoretical speculations as certainty bounds 
(see e.g. Ref.~\cite{Erler}).

I can imagine that at this point there are still those
who would like to give limits which sound probabilistical.
I hope that I have convinced them about the crucial
role of prior, and that it is not scientific to
give a confidence level  which  is 
not a `level of confidence'. In Ref.~\cite{noi} you
will find a long discussion about role and quantitative effect
of priors, about the implications of uniform
prior and so-called Jeffreys' prior, and about 
more realistic priors of experts. There, it has also been shown
that (somewhat similar to of what was said in the previous
section) it is possible to choose a prior which  
provides practically the same probabilistic result 
acceptable to all those 
who share a similar scientific prejudice. This scientific prejudice
is that of the `positive attitude of physicists'~\cite{YR}, 
according to which rational and responsible people who have 
planned, financed and run an experiment, 
consider they have some reasonable chance to observe 
something.\footnote{In some cases researchers are aware 
of having very little chance of observing anything, but 
they pursue the research to refine instrumentation and 
analysis tools
in view of some positive results in the future. 
A typical case is gravitational wave search. In this case 
it is not scientifically correct to provide probabilistic 
upper limits from the current detectors,
and the honest way to provide the 
result is that described here~\cite{PP}.
However, some could be tempted to 
use a frequentistic procedure 
which provided an `objective' upper limit 
`guaranteed' to have a 95\% coverage. 
This behaviour is irresponsible since 
these researchers are practically sure that
the true value is below the limit. 
Loredo shows in Section 3.2 of  Ref.~\cite{Loredo} 
an instructive real-live example of a 90\% C.I. 
which certainly does not contain the true value 
(the web site \cite{Loredo} contains several direct comparisons
between frequentistic versus Bayesian results.).}
It is interesting  that,
no matter how this `positive attitude' is reasonably
modelled, the final p.d.f. is, for the case of g.w. bursts
($\mu_{\mbox{ins}}=0$), 
very similar to that obtained by a uniform distribution. 
Therefore, a uniform prior could be used to provide 
some kind of conventional probabilistic upper limits, 
which could look acceptable to all those who share 
that kind of positive attitude. But, certainly, 
it is not possible to pretend that these probabilistic
conclusions can be shared by everyone. 
Note that, however, this  
idea cannot be applied in a straightforward way 
in case $\mu_{\mbox{ins}}=\infty$, as can be easily
understood. In this case one can work on a sensible 
conjugate variable (see next section)
which has the asymptotic insensitivity limit at 0,
as happens, for example, 
with $\epsilon/\Lambda^2$ in the case of a search 
for contact interaction, as initially proposed in 
Refs.~\cite{Moriond90,CELLO} and still currently done 
(see e.g. Ref.~\cite{zeus}). Ref.~\cite{Moriond90}
contains also the basic idea of using a sensitivity bound, 
though formulated differently in terms of `resolution power cut-off'.

\section{OPEN VERSUS CLOSED LIKELIHOOD}\label{sec:routine}

Although the extended discussion on priors has been
addressed elsewhere~\cite{noi},
Figs. \ref{fig:rbur}, \ref{fig:R_higgs} and \ref{fig:R_zeus} show clearly 
the reason that frontier measurements are 
crucially dependent on priors:  the likelihood
only vanishes on one side (let us call these 
measurements  `open likelihood').
In other cases the likelihood 
goes to zero in both sides (closed likelihood). 
Normal routine measurements belong to the second class,
and usually they are characterized by a narrow likelihood,
meaning high precision. Most particle physics 
measurements belong to the class of closed priors. 
I am quite convinced that 
the two classes should 
be treated routinely differently. This  does not mean 
recovering frequentistic `flip-flop'
(see Ref.~\cite{FC} and references therein), 
but recognizing the 
qualitative, not just quantitative, difference 
between the two cases, and treating them differently. 

 When the 
likelihood is closed, the sensitivity on the choice of prior
is much reduced, and a probabilistic result 
can be easily given. The subcase better understood 
is when the likelihood is very narrow. 
Any reasonable prior which models 
the knowledge of the expert interested in the inference 
is practically constant in the narrow range 
around the maximum of the likelihood. Therefore,
we get the same result obtained by a uniform prior. 
However, when the likelihood is not so narrow, 
there could still be some dependence on the metric used. 
Again, this problem
has no solution if one considers inference as a 
mathematical game~\cite{priors}. Things are less problematic
if one uses physics intuition and experience. 
The idea is to use 
a uniform prior on the quantity which is `naturally measured'
by the experiment. 
This might look like an arbitrary concept, but is in fact
an idea to which experienced physicists are accustomed.
For example, we say that `a tracking devise measures $1/p$',
`radiative corrections measure $\log(M_H)$',
`a neutrino mass experiment is sensitive to $m^2$',
and so on. 
We can see that our intuitive idea of `the quantity really measured'
is related to the quantity which has a linear dependence  
on the observation(s). When this is the case, 
random (Brownian) effects occurring during the process of measurement
tend to produce a roughly Gaussian distribution of observations. 
In other words, we are 
dealing with a roughly Gaussian likelihood. So, a way
to state the natural measured quantity  
is to refer to the quantity for which the likelihood is 
roughly Gaussian. This is the reason
why we are used do least-square fits choosing the variable in
which the $\chi^2$ is parabolic (i.e. the likelihood is normal)
and then interpret the result 
as probability of the true value. 
In conclusion, having to give 
a suggestion, I would recommend  continuing with the tradition
of considering natural the quantity which gives a 
 roughly normal likelihood. For example, this was the original
motivation to propose $\epsilon/\Lambda^2$ to report 
compositeness results~\cite{Moriond90}.

This  uniform-prior/Gaussian-likelihood 
duality goes back to Gauss himself~\cite{Gauss}.
 In fact, he derived his famous distribution to solve 
an inferential problem using what we call nowadays 
the Bayesian approach. Indeed, he assumed  
a uniform prior for the true value (as Laplace did) and searched for the 
analytical form of the likelihood such as to
give a posterior p.d.f. with most probable\footnote{Note
that also speaking about the most probable value is close
to our intuition, although all values have zero probability. 
See comments in Section 4.1.2 of Ref.~\cite{YR}.} value 
equal to the arithmetic average of the observation. 
 The resulting function was  \ldots the Gaussian.

\begin{figure}
\begin{center}
\begin{tabular}{cc}
\epsfig{file=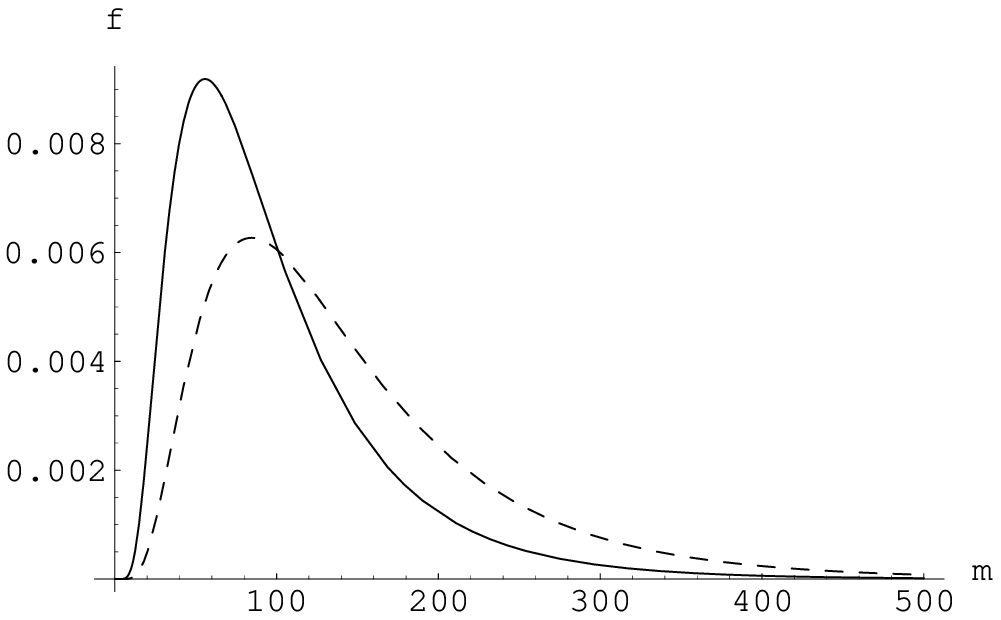,width=0.48\linewidth,clip=} &
\epsfig{file=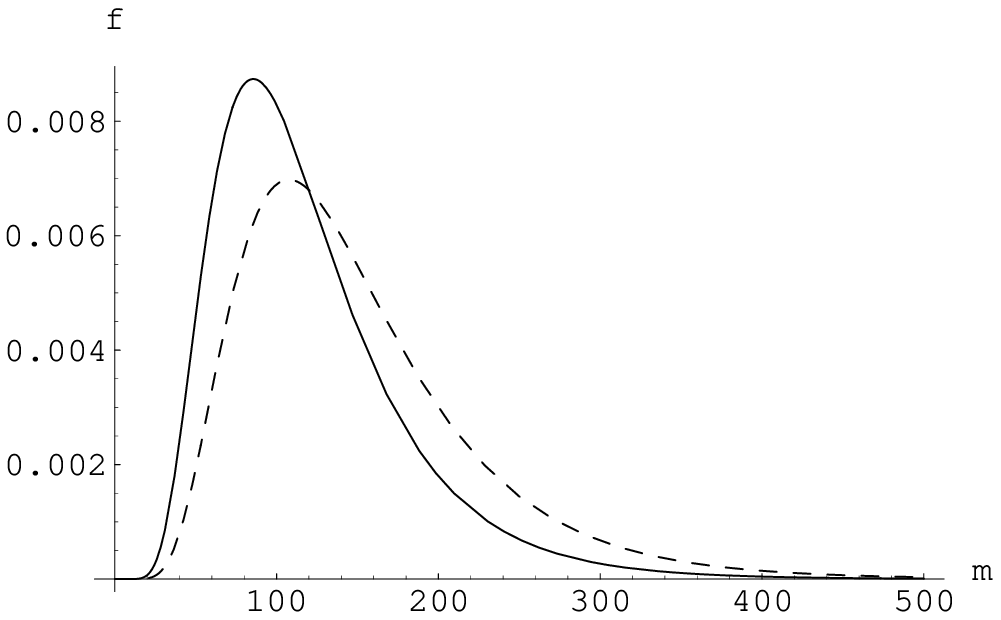,width=0.48\linewidth,clip=}  
\end{tabular}
\end{center}
\vspace{-5cm}\hspace{+3.5cm}$(\Delta \alpha)=0.02804(65)$
\hspace{+5.0cm}$(\Delta \alpha)=0.02770(65)$

\vspace{4.2cm}

\caption{\small Sensitivity analysis exercise from  the indirect 
Higgs mass determination of Ref.~\cite{Higgs_clw}. 
Solid lines and dashed lines are obtained with priors
uniform in $\log(m_H)$ and $m_H$, respectively.}
\label{fig:sens}
\end{figure}

When there is not an agreement about the natural
quantity one can make a sensitivity analysis of the result,
as in the exercise of Fig.~\ref{fig:sens},
based on Ref.~\cite{Higgs_clw}. 
If one chooses a prior flat 
in $m_H$, rather than in $\log(m_H)$, the p.d.f.'s given 
by the continuous curves change into the dashed ones. 
Expected value and standard deviation of the distributions
(last digits in parentheses) 
change as follows. For $(\Delta \alpha)=0.02804(65)$,
 $M_H=0.10(7)$ TeV becomes $M_H=0.14(9)$ TeV, 
while for $(\Delta \alpha)=0.02770(65)$
$M_H=0.12(6)$ TeV becomes $M_H=0.15(7)$ TeV. 
Although this is just an academic exercise, since it is rather well
accepted that radiative corrections measure $\log(M_H)$,
Fig.~\ref{fig:sens} and the above digits show that 
the result is indeed rather stable: 
$0.15(9)\approx 0.10(7)$ and $0.15(7)\approx 0.12(6)$,
though perhaps some numerologically-oriented colleague would disagree.

If a case is really controversial, one can still
show the likelihood. But it is important to understand that
a likelihood is not yet the probabilistic result 
we physicists want. If only the likelihood is published, the risk
it is too high  that 
it will considered anyway and somehow as a probabilistic 
result, as happens now in practice. For this reason, 
I think that, at least in the rather simple case of closed likelihood,
those who perform the research should take their responsibility
and assess expected value and standard deviation that 
they really believe, plus other information in the case
of a strongly non-Gaussian distribution~\cite{Higgs,Higgs_clw,asym}. 
I do not think that, in most applications, this subjective 
ingredient is
more relevant than the many other subjective 
choices made during the experimental activity and that we have
accept anyhow. In my opinion, adhering strictly to the point of view
that one should refrain totally from giving probabilistic results
because of the idealistic principle of avoiding the contribution
of personal priors will halt research. We always rely on 
somebody else's 
priors and consult experts. Only a perfect idiot has no prior, 
and he is not the best person to consult. 

\section{OVERALL CONSISTENCY OF DATA}\label{sec:consistency}

One of the reasons  for confusion with confidence levels 
is that the 
symbol `C.L.' is  not only used in conjunction with 
confidence intervals, 
but also associated with results of a fits, in the sense
of statistical significance (see e.g. Ref.~\cite{PDG}). 
As I have commented elsewhere~\cite{maxent98,YR}, the problem
coming from the misinterpretation of confidence levels 
are much more severe than than what happens 
considering confidence intervals 
probabilistic intervals. Sentences like
``since the fit to the data yields  
a 1\% C.L., the  theory has a 1\% chance of being correct''
are rather frequent. 
Here I would like only to touch some points which I consider important. 

Take the $\chi^2$, certainly the most used test variable 
in particle physics.
As most people know from the theory, 
and some from having had bad experiences 
in  practice, the $\chi^2$ is not what statisticians call a `sufficient
statistics'.  This is the reason why, if we see a discrepancy
in the data, but the $\chi^2$ doesn't say so, other pieces of
magic are tried, like changing the region in which the $\chi^2$ 
is applied, or using a `run test', Kolmogorov test, 
and so on\footnote{Everybody has
experienced endless discussions on what I call all-together
$\chi^2$-ology, to decide if there is some effect.}
(but, ``if I have to draw  conclusions from 
a test with a Russian name, it is better I redo the experiments'',
somebody once said). 
My recommendation is to give always a look
at the data, since the eye of the expert is in most simple 
(i.e. low-dimensional) cases better that automatic tests (it is
also not a mystery that tests are done with the hope they will prove 
what one sees\ldots). 

I think that $\chi^2$,
as other variables,
can be used {cum grano salis}\footnote{See Section 8.8 
of Ref.~\cite{YR} for a discussion
about why frequentistic tests `often work'.} 
to spot a possible problem of the experiment, 
or hints of new physics, which one 
certainly has to investigate.
What is important is to be careful 
before drawing conclusions only from the crude result of the test.
I also find it important to start calling things by 
their name  in our community too 
and call `P-value' the number resulting from the test, 
as is currently done in modern books of statistics
(see e.g.  \cite{Cowan}). 
It is recognized by statisticians that  P-values also tend to be 
misunderstood~\cite{Science,BB}, but at least they have a more precise
meaning\cite{Scherwish} than our ubiquitous C.L.'s.  

The next step is what to do when, no matter how, one has strong doubts
about some anomaly. Good experimentalists know  
their job well: check everything possible, calibrate the components,
make special runs and Monte Carlo studies, or even repeat 
the experiment, if possible. 
It is also well understood that it is not easy to
decide when to stop making studies and 
applying corrections. The risk to influencing  a result
is always present. I don't think there is any general advice that  
that can be given. Good results come from 
well-trained (prior knowledge!) honest physicists
(and who are not particularly unlucky\ldots).

A different problem is what to do when we have 
to use someone else's results, about which we do not have 
inside knowledge, for example when we make global fits. 
Also in this case I mistrust 
automatic prescriptions~\cite{PDG}.
In my opinion, when the data points appear 
somewhat inconsistent with each other (no matter how one has formed
this opinion) one has to try to model one's scepticism. Also in this case,
the Bayesian approach offers valid help\cite{Press,Dose}.
In fact, since one can 
assign probability to every piece of information which is not 
considered certain, it is possible to build a so-called 
probabilistic network~\cite{BN}, 
or Bayesian network, to model the problem
and find the most likely solution, given well-stated assumptions.
A first application 
of this reasoning in particle physics data
(though the problem was too trivial to build up
a probabilistic network representation)
is given in Ref.~\cite{sceptical},
 based on an improved
version of Ref.~\cite{Dose}.

\section{CONCLUSION}

So, {\it what is the problem?} In my opinion the root of the 
problem is the frequentistic intrusion
into the natural approach initially followed by `classical' physicists
and mathematicians (Laplace, Gauss, etc.) to solve 
inferential problems. As a consequence,
we have been taught to 
make inferences using statistical methods which were not 
conceived  for that purpose, as insightfully illustrated 
by a professional statistician at the workshop\cite{Clifford}. 
It is a matter of fact that the results of these 
methods are never intuitive (though we force the `correct' 
interpretation using out intuition~\cite{maxent98}),  
and fail any time the 
problem is not trivial. The problem of the limits in 
`difficult cases' is particularly evident, because these
methods fail~\cite{Zech1}. But I would like to remember 
that also in simpler routine problems, like uncertainty propagation
and treatment of systematic effects, conventional
statistics do not provide consistent
methods, but only a prescription which we are supposed 
to obey. 

{\it What is the solution?} As well expressed in Ref.~\cite{WWF},
sometimes we cannot solve a problem because we are not 
able to make a real change, and we are trapped in a kind of
logical maze made by many solutions, which are not the solution. 
Ref.~\cite{WWF} talks explicitly of non-solutions forming 
a kind of group structure. We rotate
inside the group, but we cannot solve the problem until we  
break out of the group. I consider the many attempts to solve the problem
of the confidence limit inside the frequentistic framework as just
some of the possible group rotations. 
Therefore the only possible solution I see is to get rid 
of frequentistic intrusion in the natural 
physicist's probabilistic reasoning. This way out, which 
takes us back the `classicals', is offered by the statistical 
theory called Bayesian, a bad name
that gives the impression of a religious sect to which we 
have to become converted (but physicists will never be Bayesian,
as they are not Fermian or Einsteinian~\cite{maxent98} -- why
should they be Neymanian or Fisherian?). 
I consider the name Bayesian to be temporary and
just in contrast to
`conventional'. 

I imagine, and have experienced,
much resistance to this change due to educational, 
psychological and cultural reasons (not forgetting 
the sociological ones, usually the hardest ones to remove). 
For example, a good cultural reason is that we consider,
in good faith, a statistical theory on the same footing 
as a physical theory. We are used to a well-established 
physical theory being better than the previous one.
This is
not the case of the so-called classical statistical theory,
and this is the reason why an increasing number of 
statisticians and scientists ~\cite{Science} have restarted from 
the basic ideas of 200 years ago, complemented with 
modern ideas and computing 
capability~\cite{BN,deFinetti,Jaynes,Sivia,Loredo,Bayesian}.
Also in physics things are moving, and there are many now 
who oscillate between the two approaches, 
saying that both have  good and bad features.
The reason I am rather radical is because  I do not
think we, as physicists, should care only about numbers, 
but also about their meaning: 
25 is not approximatively equal to 26, 
if 25 is a mass in kilogrammes
and 26 a length in metres. 
In the Bayesian approach I am confident of what numbers 
mean at every step, and how to go further. 

I also understand that sometimes things are not so 
obvious or so highly intersubjective, as an
anti-Bayesian joke says:
``there is one obvious possible way to do things, 
it's just that they can't agree on it.'' 
I don't consider this a problem. In general, it is just 
due to our human condition when faced with the unknown
and to the fact that (fortunately!) we do not have
an identical status of information. 
But sometimes the reason is more trivial, that is
we have not worked together enough on common problems. 
Anyway, given the choice between  a set of prescriptions 
which gives an exact (`objective')
value of something which has no meaning, 
and a framework which gives a rough value
of something which has a precise meaning,
I have no doubt which to choose. 


Coming, finally, to the specific topic of the workshop,
things become quite easy, once we have understood why
an objective inference cannot exist, but an `objective' (i.e.
logical) inferential framework does.
\begin{itemize}
\item
In the case of open likelihood, priors become
crucial. The likelihood (or the 
${\cal R}$-function) should always be reported, 
and a non-probabilistic 
sensitivity bound should be given to summarize the 
negative search with just a number. 
A conventional probabilistic result can be provided 
using a uniform prior in the most natural quantity.
Reporting the results with the ${\cal R}$-function 
satisfies the desiderata expressed in this paper.
\item
In the case of closed likelihood, a uniform prior 
in the natural quantity provides probabilistic results which
can be easily shared by the experts of the field. 
\end{itemize}
As a final remark, 
I would like to recommend calling things by their name,
if this name has a precise meaning. In particular: 
sensitivity bound if it is just a sensitivity bound,
without probabilistic meaning; and 
such and such percent probabilistic limit, if it really
expresses the confidence of the person(s) who assesses it.
As a consequence, I would propose not to
talk any longer about `confidence interval' and `confidence level',
and to abandon the abbreviation `C.L.'. So, although it might 
look paradoxical, I think that {\it the} solution
to the problem of confidence limits begins with  
removing the expression itself. 



\begin{thebibliography}{99}

%
\bibitem{cl_papers}
P. Janot and F. Le Diberder, 
 {\it Combining `limits'}, 
CERN--PPE--97--053, May 1997, \\
%
A. Favara and M. Pieri, {\it Confidence level estimation and 
analysis optimization}, internal report DFF--278--4--1997 (University 
of Florence), {\tt hep-ex/9706016}; \\
%
B.A. Berg and I-O Stamatescu, {\it Neural networks and confidence
limit estimates}, FSU--SCRI--98--08 (Florida State University), January 1988.
%
P. Janot and F. Le Diberder, 
{\it Optimally combined confidence limits},
Nucl. Instrum. Methods, {\bf A411} (1998) 449; \\
%
D. Silverman, {\it Joint Bayesian treatment of Poisson and Gaussian 
experiments in chi-squared statistics}, 
U.C. Irvine TR--98--15, October 1998, {\tt physics/9808004}; \\
%
C. Giunti, {\it A new ordering principle for the classical statistical
analysis of Poisson processes with background},
Phys. Rev. {\bf D59} (1999) 053001;\\
%
C. Giunti, {\it Statistical interpretation of the null result of 
the KARMEN 2 experiment}, internal report 
DFTT--50--98 (University of Turin), {\tt hep-ph/9808405};\\
%
S. Jim and P. McNamara, {\it The signal estimator limit setting 
method}, {\tt physics/9812030};\\
%
B.P. Roe and  M.B. Woodroofe, {\it Improved probability method 
for estimating signal in the presence of background}, 
Phys. Rev. {\bf D60} (1999) 053009;\\
%
C. Giunti, {\it Treatment of the background error in the statistical
analysis of Poisson processes},  
Phys. Rev. {\bf D59} (1999) 113009;\\
%
T. Junk, {\it Confidence level computation for combining searches
with small statistics},
 Nucl. Instrum. Methods {\bf A434} (1999) 435. 
%
S.J. Yellin, {\it A comparison of the LSND and KARMEN 
$\overline{\nu}$ 
oscillation experiments},
Proc.  COSMO 98, Monterey, CA, 15--20 November 1998, 
 {\tt hep-ex/9902012};\\
%
S. Geer, {\it A method to calculate limits in absence 
of a reliable background subtraction}, 
Fermilab--TM--2065, March 1999; \\
%
I. Narsky, {\it Estimation of upper limits using a Poisson statistics},
{\tt hep-ex/9904025}, April 1999;\\
%
H. Hu and J. Nielsen, {\it Analytic confidence level calculations 
using the likelihood ratio and Fourier transform}, 
{\tt physics/9906010}, June 1999;\\
%
O. Helene, {\it Expected coverage of Bayesian and classical 
intervals for a small number of events}, 
Phys. Rev. {\bf D60} (1999) 037901;\\
%
J.A. Aguilar-Saavedra, 
{\it Computation of confidence intervals for Poisson processes},
UG--FT--108/99, November 1999, {\tt hep-ex/9911024};\\
%
M. Mandelkern and J. Schultz,
{\it The statistical analysis of Gaussian and Poisson signals
near physical boundaries}, {\tt v2}, December 1999, 
{\tt hep-ex/9910041};\\
%
J. Bouchez, 
{\it Confidence belts on bounded parameters},  
January 2000, {\tt hep-ex/0001036};\\
%
C. Giunti, M. Laveder, 
{\it The statistical and physical significance of confidence intervals},
{\tt hep-ex/0002020}
%
\bibitem{FC}
G.J. Feldman and R.D. Cousins, {\it Unified approach to the classical 
statistical analysis of small signal}, 
Phys. Rev. {\bf D57} (1998) 3873.
%
\bibitem{LEP}
P. Bock et al. (ALEPH, DELPHI, L3 and OPAL Collaborations),
{\it Lower bound for the standard model Higgs boson mass from combining
the results of the four LEP experiments}, 
CERN--EP/98--046, April 1998, and references therein.
%
\bibitem{PDG}
C. Caso et al., {\it Review of particle physics},
Eur. Phys. J. {\bf C3} (1998) 1 ({\tt http://pdg.lbl.gov}). 
%
\bibitem{Zech}
G. Zech, {\it Objections to the unified approach to the 
computation of classical confidence limits}, 
{\tt physics/9809035}.
%
\bibitem{Ciampolillo}
S. Ciampolillo, {\it Small signal with background: objective
confidence intervals and regions for physical parameters from 
the principle of maximum likelihood}, 
Il Nuovo Cimento {\bf 111} (1998) 1415. 
%
\bibitem{ci}
 G. D'Agostini, {\it Contact interaction scale
from deep-inelastic scattering events -- what do the data teach us?},
ZEUS note 98--079, November 1998.
%
\bibitem{Higgs} 
G. D'Agostini and G. Degrassi,
{\it Constraints on the Higgs boson mass from direct searches and 
precision measurements},  Eur. Phys. J. {\bf C10} (1999) 663.
%
\bibitem{Eitel}
K. Eitel, {\it Compatibility analysis of the LSND evidence 
and the KARMEN exclusion for 
$\overline{\nu}_\mu \rightarrow \overline{\nu}_e$ oscillations},
{\tt hep-ex/9909036}. 
%
\bibitem{noi}
P. Astone and G. D'Agostini, 
{\it Inferring the intensity of Poisson processes at the limit 
of the detector sensitivity (with a case study on gravitational 
wave burst search)}, CERN--EP/99--126, August 1999, {\tt hep-ex/9909047}.
%
%
\bibitem{Punzi}
G. Punzi, {\it A stronger classical definition of confidence limits},
December 1999, {\tt hep-ex/9912048}.
%
\bibitem{CLW}
Workshop on `Confidence Limits', CERN, Geneva, 17--18 January 2000,\\
{\tt http://\-www.\-cern.\-ch/\-CERN/\-Divisions/\-EP/Events/CLW/}.
%
\bibitem{Fisher} 
R.A. Fisher, {\it Statistical methods and scientific induction},
J. Royal Stat. Soc. {\bf B 17} (1955) 69.
%
\bibitem{Hald}
A. Hald, {\sl A history of mathematical statistics from 1750 to 1930}
(John Wiley \& Sons, 1998).
%
\bibitem{maxent98} 
G. D'Agostini, {\it Bayesian reasoning versus conventional statistics
in high energy physics}, 
Proc.  XVIII International Workshop 
on Maximum Entropy and Bayesian Methods, Garching, Germany, July 1998
(Kluwer Academic, 1999), pp. 157--170, 
 {\tt physics/9811046}.
%
\bibitem{Frodesen}
A.G. Frodesen, O. Skjeggestad and H. Tofte,
{\sl Probability and statistics in particle physics}, 
Columbia University, New York, 1979.
%
\bibitem{Eadie}
W.T. Eadie, D. Drijard, F.E. James, M. Roos and B. Sadoulet, 
{\sl Statistical methods in experimental physics}
(North Holland, Amsterdam, 1971). 
%
\bibitem{Science}
D. Malakoff, {\it Bayes offers a `new' way to make sense of numbers}, 
Science {\bf 286}, 19 November 1999, 1460--1464. 
%
\bibitem{YR}
G. D'Agostini, {\it Probabilistic reasoning in HEP -
principles and applications},
Report CERN 99--03, July 1999, also available at the author's URL,
together with FAQs.
%
\bibitem{ajp}
G. D'Agostini, {\it Teaching statistics in the physics 
curriculum. Clarifying and unifying role of subjective probability}, 
Am. J. Phys. {\bf 67} (1999) 1260.
%
\bibitem{Jaynes}
E. T. Jaynes, {\sl Probability theory: the logic of science}, 
posthumous book in preparation, online version at 
{\tt http://bayes.wustl.edu/etj/prob.html}
%
\bibitem{priors}
G. D'Agostini, {\it Overcoming priors anxiety}, 
{\tt physics/9906048}, June 1999.
%
\bibitem{ISO}
International Organization for Standardization (ISO),
{\it Guide to the expression of uncertainty in measurement}
(ISO, Geneva, 1993).
%
\bibitem{Barlow}
R.J. Barlow, {\sl Statistics} (John Wiley \& Sons, 1989).
%
\bibitem{Glymour}
C. Glymour, {\sl Thinking things through: an introduction
to philosophical issues and achievements} (MIT Press, 1997). 
%

\bibitem{deFinetti}
B. de Finetti, {\sl Theory of probability}, 1970 (J. Wiley and Sons, 1974). 
%
\bibitem{Poincare}
H. Poincar\'e, {\sl Science and hypothesis}, 1905 
(Dover Publications, 1952).
%
\bibitem{Lavine}
M. Lavine and M.J. Schervich, {\it Bayes factors: 
what they are and what they are not}, Am. Stat. {\bf 53} (1999) 119.
%
\bibitem{clw_qm}
G. D'Agostini,
{\it Quantum mechanics and interpretation of probability  
         (with comments on confidence intervals)}, 
Contribution to this workshop (see discussion session). 
%
\bibitem{Feynman}
R. Feynman, {\sl The character of the physical law}, 1965 
(MIT Press, 1967). 
%
\bibitem{Sivia}
D.S. Sivia, {\sl Data analysis -- a Bayesian tutorial} (Oxford, 1997).
%
\bibitem{Keynes}
J.M. Keynes, {\it A tract on monetary reform}, 1927.
%
\bibitem{Erler}
J. Erler and P. Langacker,
{\it Status of the Standard Model},
Proc. 5th International WEIN Symposium, Santa Fe, NM, USA,
14--21 June 1998, {\tt hep-ph/\-9809352}.
%
\bibitem{Higgs_clw} 
G. D'Agostini and G. Degrassi,
{\it Constraining the Higgs boson mass through the combination
of direct search and precision measurement results}, 
{\tt hep-ph/\-0001269}.
%
\bibitem{BN}
J. Pearl, {\sl Probabilistic reasoning in intelligent systems:
networks of plausible inference} (Morgan Kaufmann Publishers, 1988);
F.V. Jensen, {\sl An introduction to Bayesian networks}
(Springer, 1996);
R.G. Cowell, A.P. Dawid, S.L. Laurintzen and D.J. Spiegelhalter,
{\sl Probabilistic networks and expert systems} (Springer, 1999);
see also {\tt http://www.auai.org/} and  {\tt http://www.hugin.dk}.  
%
\bibitem{Pia}
P. Astone and G. Pizzella, {\it Upper Limits in the Case That Zero
 Events are Observed: An Intuitive Solution to the Background
Dependence Puzzle}, contribution to this workshop, 
{\tt hep-ex/0002028}.

\bibitem{asym}
G. D'Agostini and M. Raso,
{\it Uncertainties due to imperfect knowledge of systematic 
effects: general considerations and approximate formulae}, 
CERN--EP/2000--026, February 2000, \\ 
{\tt hep-ex/0002056}.

%
\bibitem{zeus}
ZEUS Collaboration, J. Breitweg  et al., 
{\it Search for contact interactions
in deep-inelastic $e^+p\rightarrow e^+X$ scattering at HERA},
DESY 99--058, May 1999, 
{\tt hep-ex/\-9905039}.
%
\bibitem{Doucet}
M. Doucet, {\it Bayesian presentation of neutrino oscillation results},
contribution to this workshop. 
%
\bibitem{PP}
P. Astone and  G. Pizzella, {\it On upper limits for 
gravitational radiation}, January 2000, \\ {\tt gr-qc/0001035}. 
%
\bibitem{Loredo}
T.J. Loredo, {\sl The promise of Bayesian inference 
for astrophysics}, 
{\tt http://\-astrosun.\-tn.\-cornell.edu/\tt staff/loredo/bayes/tjl.html}
(this web site contains  also other interesting tutorials, papers
and links). 
%
\bibitem{Moriond90}
G. D'Agostini, {\it Limits on electron compositeness from
the Bhabha scattering at PEP and PETRA}, 
Proc. XXVth Rencontres de Moriond, Les Arcs, France,  4--11 March, 1990
(also DESY--90--093). 
%
\bibitem{CELLO}
CELLO Collaboration, H.J. Behrend et al.,
{\it Search for substructures of leptons and quarks with
the CELLO detector},
 Z. Phys. {\bf C51} (1991) 149.
%
\bibitem{Gauss}
C.F. Gauss, {\sl Theoria motus corporum coelestium in sectionibus 
conicis solem ambientum}, Hamburg 1809, n.i 172--179; reprinted
in Werke, Vol. 7 (Gota, G\"ottingen, 1871), pp 225--234
(see also Ref.~\cite{Hald})
%
\bibitem{Cowan}
G. Cowan, {\sl Statistical data analysis} (Clarendon Press, Oxford, 1988).
%
\bibitem{BB}
J.O. Berger and D.A. Berry,
{\it Statistical analysis and the illusion of objectivity},
American Scientist {\bf 76} (1988) 159. 
%
\bibitem{Scherwish}
M.J. Scherwish, 
{\it P values: what they are and what they are not}, 
Am. Stat. {\bf 50} (1996) 203. 

\bibitem{Press}
W.H. Press,
{\it Understanding data better with Bayesian and global 
statistical methods}, \\ {\tt astro-ph/9604126}.

\bibitem{Dose}
V. Dose and W. von Linden, {\it Outlier tolerant parameter
estimation}, Proc. XVIII Workshop on Maximum Entropy and Bayesian
Methods, Garching, Germany, July 1998, pp. 47-56. 

\bibitem{sceptical}
G. D'Agostini, {\it Sceptical combination of experimental results:
general considerations and application to $\epsilon^\prime/\epsilon$},
CERN--EP/99--139, October 1999, {\tt hep-ex/\-9910036}.

\bibitem{Clifford}
P. Clifford, {\it Interval estimation as seen from the 
world of mathematical statistics}, contribution to this workshop. 

\bibitem{Zech1}
G. Zech, {\it Classical and Bayesian confidence limits},
paper in preparation.

\bibitem{WWF}
P. Watzlawick, J.H. Weakland and R. Fisch,
{\sl Change: principles of problem
formation and problem resolution}, W.W. Norton, New York, 1974.

\bibitem{Bayesian}
H. Jeffreys, {\sl Theory of probability} (Oxford, 1961);
J.M. Bernardo and A.F.M. Smith, {\sl Bayesian theory} 
(John Wiley and Sons, 1994); 
A. O'Hagan, {\sl Bayesian inference}, Vol. 2B of Kendall's
advanced theory of statistics (Halsted Press), 1994;
B. Buck and V.A. Macaulay, {\sl Maximum entropy in action} (Oxford, 1991);
A. Gelman, J.B. Carlin, H.S. Stern and D.B. Rubin, {\sl Bayesian 
data analysis} (Chapman \& Hall, 1995); 
\end{thebibliography}
\end{document}